# Role of intergranular silver in modulating the aperiodicity in disordered Josephson junction arrays: impact on relaxation of intergranular critical state

A. Pandey, Dipten Bhattacharya *, R.G. Sharma

*Cryogenic Division, National Physical Laboratory, Dr. K.S. Krishnan Road, New Delhi 110 012, India*

## Abstract

Relaxation of the intergranular critical state has been observed at a very low applied magnetic field (10–50 Oe) over a temperature regime of 20–77 K in bulk polycrystalline $YBa_2Cu_3O_{7-x}$ (YBCO) and $Bi_{1.75}Pb_{0.35}Sr_2Ca_2Cu_3O_{10+x}$ (BPSCCO) samples. In such a disordered Josephson junction network, the relaxation is slower than the intragranular relaxation and, hence, it yields higher flux pinning energy $U_0$ than the corresponding intragranular values. Silver addition seems to have given rise to much uniformity in the grain boundary characteristics which results in sharp drop in the flux pinning energy since it depends on the variation of the junction coupling energy $E_J$ across the network. While $U_0 \approx 0.5$ and 0.55 eV for the parent BPSCCO and YBCO samples, respectively, in silver added (10–15 wt.%) samples the corresponding values are ~0.15 and ~0.27 eV. The distribution functions $m(U)$ for the flux pinning energy and $n(\theta)$ for the grain boundary misalignment angle $\theta$ have been evaluated from the experimentally observed patterns of magnetic relaxation and variation of the grain boundary critical fields with temperature. The distribution functions become narrower in the case of silver added samples reflecting a reduction in the degree of disorder. The variation of the effective vortex mass $m^*$ with the variation in the degree of disorder is observed. Considering the width of the superconducting transition $\Delta T_c$ as the measure of the degree of disorder (inhomogeneity), it has been shown that the transport critical current density $J_c$ follows a relationship $J_c \sim \exp(-\Delta T_c)$ while the pinning energy $U \sim \Delta T_c$. These relations may help in devising a suitable strategy for achieving the desired effect: high $J_c$ yet slower decay rate, i.e., large $U$. © 2000 Elsevier Science B.V. All rights reserved.

*PACS:* 74.80.Bj; 74.60.Ge; 74.60.Jg; 72.80.Ng

*Keywords:* Magnetic relaxation; Josephson junction array; Flux pinning; Critical current density; Transition width

## 1. Introduction

The problem of vortex transport and localization continues to be one of the mainstays of the research activities in high-$T_c$ superconducting systems. Fascinating physics ranging from critical current vs. field patterns ([1,2] and references therein) to soliton dynamics [3] and even the problem of fractional statistics [4] could be studied with vortices in confined geometries, like single or array of Josephson junctions of such high-$T_c$ systems. Vortex transport in such systems also simu-lates charge conduction in periodic or aperiodic medium in suitable limit as well as gives rise to relaxation in the magnetization, drop in critical current, noise in any electronic device fabricated out of superconducting system, etc. Pinning of vortices, therefore, assumes importance as it leads to dissipation free flow of supercurrent and noise-free performance of the devices. Flux pinning in Josephson junction array results from (i) the discreteness of the array with fluctuation in the Josephson phase linkage [5], (ii) fluctuation in the phase difference due to small inductive current loops in the junction [6] and (iii) variation in the junction coupling energy $E_J$ ($\sim hI_c/2e$, $I_c$ is the critical current of the junction) across the network due to inhomogeneities or defects [7].

* Corresponding author. Tel.: +91-11-5787161, ext. 2385; fax: +91-11-5852678.
E-mail address: dipten@csnpl.ren.nic.in (D. Bhattacharya).

Role of bulk pinning in large Josephson junction (where either the length or width of the junction is larger than the Josephson penetration depth $\lambda_J$ has been studied for different types of defects (periodic or random with varying sizes) by Fehrenbacher et al. [8] through exact solution of the sine-Gordon equation. Experimentally also, the improvement of the junction critical current at commensurate flux concentration has been observed [9,10]. Such observations prompt one to study the flux pinning as well as critical current in a disordered junction array. All these features are quite relevant in describing the flux pinning and current transport in granular high-$T_c$ systems. In fact, bulk polycrystalline high-$T_c$ systems provide opportunities to study these features.

Earlier studies [11–13],[1] in this context, have highlighted that though the bulk high-$T_c$ systems offer higher Josephson vortex pinning energy as a result of large scale inhomogeneity (and consequent variation in the junction coupling energy $E_J$), the net area of high current carrying cross-section is small in such systems which, in turn, leads to smaller transport critical current. Uniform grain boundary characteristics across the sample (achieved either through silver addition in the matrix or by better processing) can lead to a drop in flux pinning energy, rise in the net area of high current carrying cross-section and consequently high transport critical current.

Given such backdrop, we report in this paper, our systematic results of magnetic relaxation studies under low field (10–50 Oe) over a temperature range of 20–77 K in pure and 10–15 wt.% silver added bulk polycrystalline Y–Ba–Cu–O and Bi–Pb–Sr–Ca–Cu–O systems. Under such low field, the vortices are present only at the grain boundaries and, therefore, the flux pinning or vortex dynamics is related to the pinning in disordered grain boundary array. We have observed that the flux pinning energy drops by half or one third in the silver added systems which signifies a marked improvement in the uniformity in the grain boundary characteristics across the sample. We have estimated the distribution of the flux pinning energies as well as grain boundary misalignment angles across the bulk of the systems from the magnetic relaxation patterns and variation in the grain boundary critical fields with temperature. The distribution in the case of silver added systems is found to be narrower. We have also observed that the transport $J_c$ follows $\sim \exp[-\Delta T_c](\Delta T_c =$ superconducting transition width) pattern while the flux pinning energy $U$ follows $\sim \Delta T_c$ pattern. Therefore, it seems that in order to achieve high transport $J_c$ and high flux pinning energy $U$, an optimum degree of uniformity in the grain boundary network is to be achieved.

## 2. Experimental details

Bulk $YBa_2Cu_3O_{7-x}$ (YBCO) and $Bi_{1.75}Pb_{0.35}Sr_2Ca_2Cu_3O_{10+x}$ (BPSCCO) sintered discs were prepared for the entire study. The powder is prepared by solid state reaction and solution chemistry route. In the case of YBCO sample, HgO was used as an internal oxygen source and hence, the samples could be sintered in air. The details of the sample preparations can be found elsewhere [14]. BPSCCO powder were prepared by a technique of autoignition of citrate–nitrate gel. The details of the powder preparation is published elsewhere [15]. The starting composi-tion is selected from the optimum range outlined in Ref. [16] which is expected to maximize the high-$T_c$ 2223 phase formation. Silver is added to the calcined product of both the YBCO and BPSCCO systems in 10–15 wt.% proportion as well as in the mixture of metal nitrates in the case of BPSCCO system as silver nitrate. Microscopic characterizations, such as SEM, XRD, EDX, etc., were carried out for all these samples in order to study the microstructure, phase purity, compositional homogeneity, etc. The distribution of silver across the entire bulk is mapped using the EDX signal. The distribution is found to be reasonably uniform (Fig. 1) which leads to the development of grain boundaries with uniform characteristics.

---

[1] Earlier observations suggested small or virtually zero relaxation of magnetization in granular medium at low temperature ($\sim$4.2 K).

Detailed magnetic studies, e.g., initial magnetization, hysteresis, relaxation effect, etc. were carried out over a temperature regime of 20–77 K using a vibrating sample magnetometer (VSM, model no. DMS 1660) coupled with a cryocooler of CTI, Cryogenics Inc. Standard four probe configuration is employed for the resistivity and transport $J_c$ measurements. The steps of the magnetic studies are as follows: (i) the sample temperature is brought down from room temperature to the desired one under zero-magnetic field; (ii) initial magnetization study is carried out at that temperature to identify the lower and upper critical fields of the grain boundaries, $H_{c1j}(T)$ and $H_{c2j}(T)$; typical patterns are shown in Fig. 2. The field corresponding to the deviation in moment vs. field pattern from straight line can be considered as $H_{c1j}(T)$ and the field corresponding to the kink (from where the negative moment starts increasing again) is considered as $H_{c2j}(T)$. These points are marked as A and B in Fig. 2. (iii) The sample is recycled back to the same temperature by heating and subsequent cooling under zero field. This step is important to drive out any remnant flux (or trapped flux) from the bulk; (iv) low-field hysteresis within 10–50 Oe is studied. The field sweeping rate for both studies was 0.15–0.2 Oe/s; the field limit is chosen in such a way that the range lies within $H_{c1j}(T)$ and $H_{c2j}(T)$ at a given temperature. Within this range, the flux lines are present only at the grain boundaries while the grains remain shielded. The induced current loop scales the entire grain–grain boundary network under such a condition. For such a sweeping rate, the rate of change in magnetization is found to be $\sim 7$–$10 \times 10^{-4}$ emu/s which leads to (from London equation) the development of $\sim 10^{-8}$–$10^{-9}$ V/cm electric field. This criterion is much more stringent than the criterion ($10^{-6}$ V/cm) used in the case of transport measurements. The impact of relaxation, therefore, is severe on the magnetic properties. (v) The magnetic relaxation study is carried out by bringing down the sample temperature from above $T_c$ to the desired one under zero field. A certain amount of field (which lies within the range $H_{c1j}(T)$–$H_{c2j}(T)$ at a given temperature) is, then, set. The relaxation (i.e., the change in magnetic moment with time) is noted over a time period of $\sim 3000$ s. After completing the study at a certain temperature, the sample is recycled back to a different temperature by warming and cooling under zero field. The process is continued for studying the relaxation effect over the entire range 20–77 K. The above steps are in conformity with the steps recommended by Yeshurun et al. [17]. The problems of field inhomogeity, demagnetization, improper field sweeping rate etc. have been taken care of through meticulous scheduling of the steps and proper selection of the sweeping rate. The demagnetization factors $d$ for each of the samples have been calculated from the slopes of the initial magnetization curves and are listed in the Table 1.

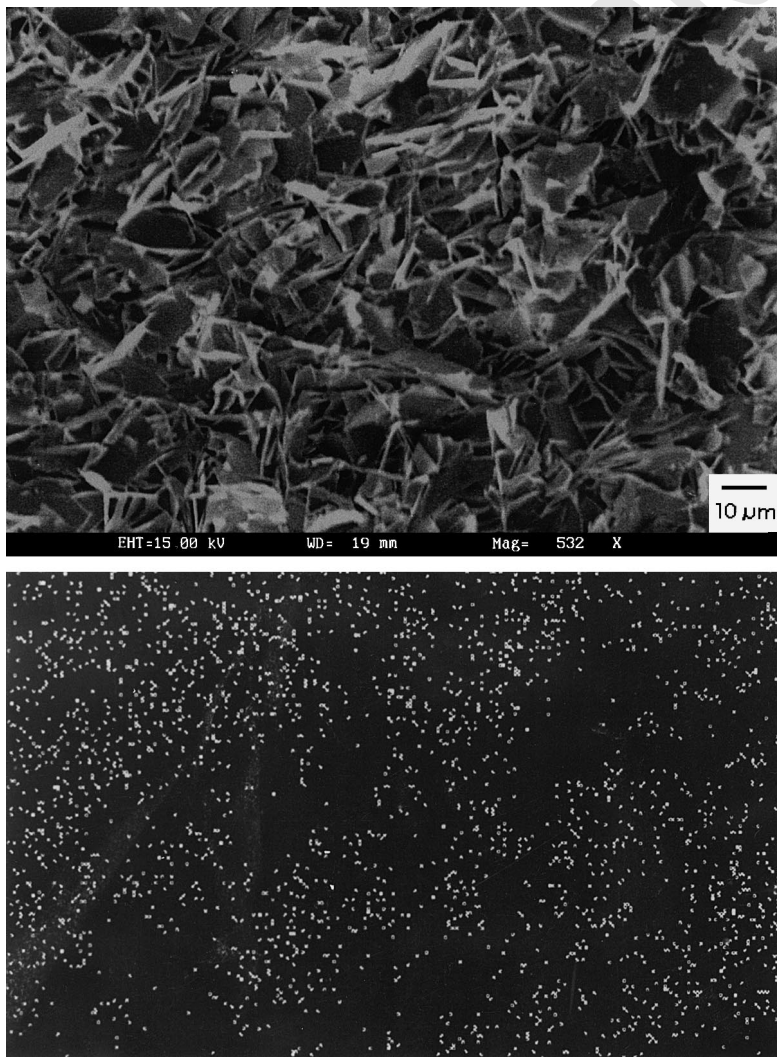

Fig. 1. A typical microstructure and the local X-ray mapping of the silver distribution across the grain–grain boundary network.

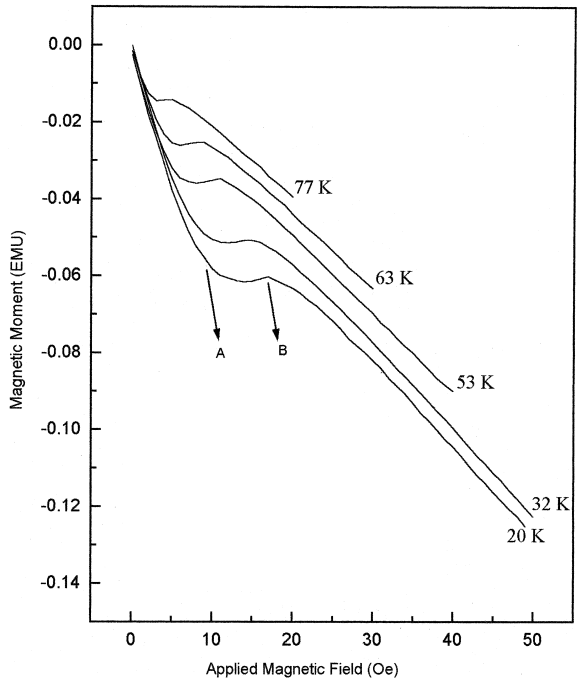

Fig. 2. A typical initial magnetization plot. The critical fields $H_{c1j}(T)$ and $H_{c2j}(T)$ are marked as A and B, respectively.

Table 1
List of few parameters relevant to the magnetic properties of the samples[a]

| Samples | $U_0$ (eV) | $n$ | $\tau$ (s) | $\gamma$ | $P_m$ | $U_m$ | $d$ |
|---|---|---|---|---|---|---|---|
| 1 | 0.50 | 0.9 | $8.2 \times 10^{-10}$ | 0.42 | 0.007 | 0.45 | 0.46 |
| 2 | 0.15 | 1.5 | $5.0 \times 10^{-10}$ | 0.45 | 0.0095 | 0.20 | 0.53 |
| 3 | 0.55 | 1.2 | $3.3 \times 10^{-12}$ | 0.40 | 0.012 | 0.58 | 0.54 |
| 4 | 0.27 | 1.65 | $2.27 \times 10^{-12}$ | 0.49 | 0.021 | 0.25 | 0.63 |

[a] Sample nos. – 1: BPSCCO, 2: BPSCCO + 10 wt.% Ag, 3: YBCO, and 4: YBCO + 15 wt.% Ag.

The difference in demagnetization factor $d$ between the parent and silver added samples leads to a variation in the demagnetizing field by ~20%. However, the observed difference in the relaxation rate $(1/M_0)dM/d\ln t$ between the parent and silver added samples is much higher and hence is not an artifact of different demagnetizing field.

## 3. Results

The magnetic relaxation results at different temperatures are shown in Fig. 3. The normalized magnetic moment $(|M|/M_0)$ vs. time $t$ ($M_0$ is the initial magnetic moment) follows approximately logarithmic trend reflecting the fact that the intergranular flux creeping also follows Anderson–Kim model. Therefore, from the relation $dM/d\ln t = -k_B T/U_0$, it is possible to calculate the flux pinning energy $U_0$. In the case of silver added samples, the relaxation pattern at 77 K crosses the patterns of lower temperature (Fig. 3b), at least, in the initial phase of the relaxation process. This is because of faster relaxation of $M$ within the time span $t_0$ (a time lag between the setting of the field and actual noting of the relaxation data) and consequent erroneous identification of $M_0$, the initial magnetization. However, over the entire time span the pattern follows the expected trend and hence the temperature dependence of the normalized relaxation rate $[(1/M_0)dM/d\ln t]$ also follows the expected behavior. The flux pinning energy is found to be ~0.5 eV in the case of parent BPSCCO sample and ~0.55 eV in the case of parent YBCO sample. The silver added samples depict a drop in $U_0$ to ~0.15 and ~0.27 eV. It is to be noted here that the entire relaxation pattern can be fitted with a 'time-independent' $U_0$ which highlights that simple Anderson–Kim flux creeping model is applicable here and $U_0$ does not follow a complex $U$–$J$ (current density) dependence. Typical temperature variation of the relaxation rate $[(1/M_0)dM/d\ln t]$ is plotted in Fig. 4. The sharp rise in $[(1/M_0)dM/d\ln t]$ in silver added samples is clearly evident. The relaxation is faster within a network of uniform superconductor–normal metal–superconductor (SNS) junctions. Using the $M(t_b, T)$ (where $t_b$ is a specific time) and $dM/d\ln t$ vs. $T$ patterns, the distribution function $m(U)$ of the flux pinning energy $U$ is calculated following an inversion scheme [18]. A typical plot of $H_{c1j}(T)$ and $H_{c2j}(T)$ is shown in Fig. 5. Using this observed pattern, it is possible to calculate the distribution function $n(\theta)$ of the grain boundary misalignment angle. The transport critical current density $J_c$ vs. the transition width $\Delta T_c$ is shown in

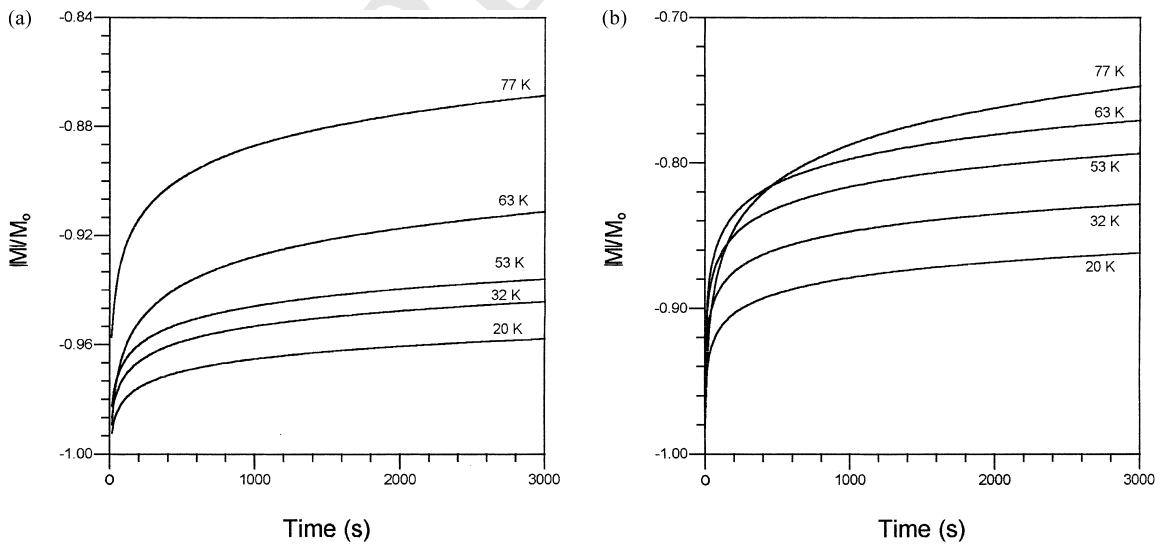

Fig. 3. Typical normalized magnetic moment vs. time pattern at different temperatures for the bulk high-$T_c$ superconductors. The patterns are observed in (a) parent and (b) 15 wt.% silver added samples.

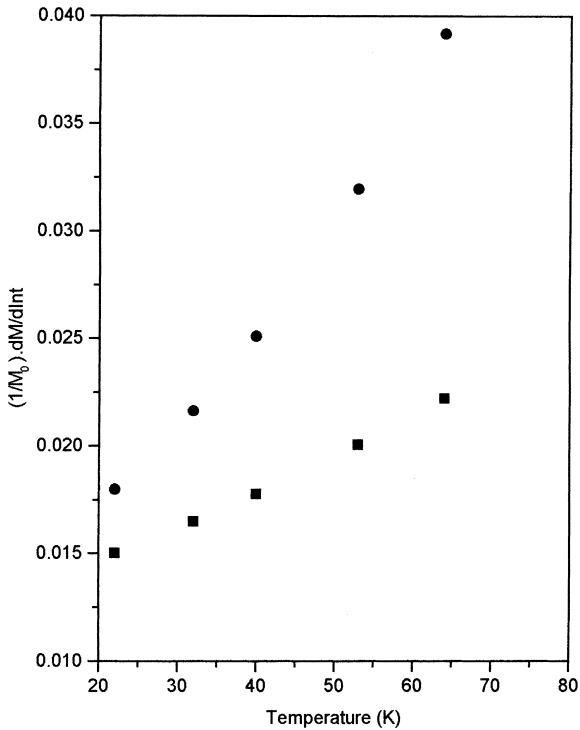

Fig. 4. Typical variation of the relaxation rate $(1/M_0)\mathrm{d}M/\mathrm{d}\ln t$ with temperature. The patterns are observed in parent (■) and 10 wt.% silver added (●) samples.

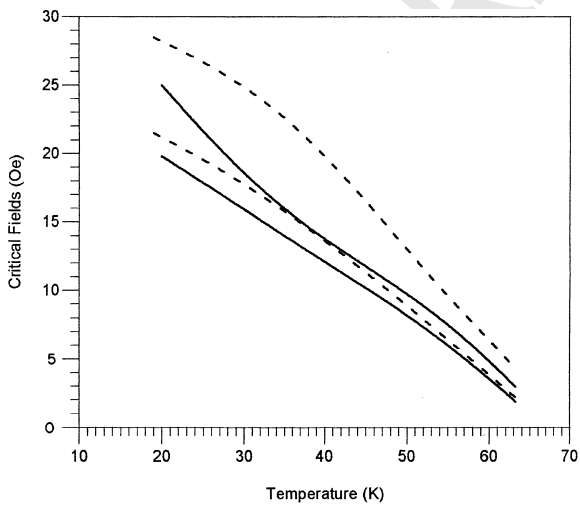

Fig. 5. A typical plot of $H_{c1j}(T)$ and $H_{c2j}(T)$ vs. temperature ($T$) for the pure sample (—) and the silver added sample (- - -).

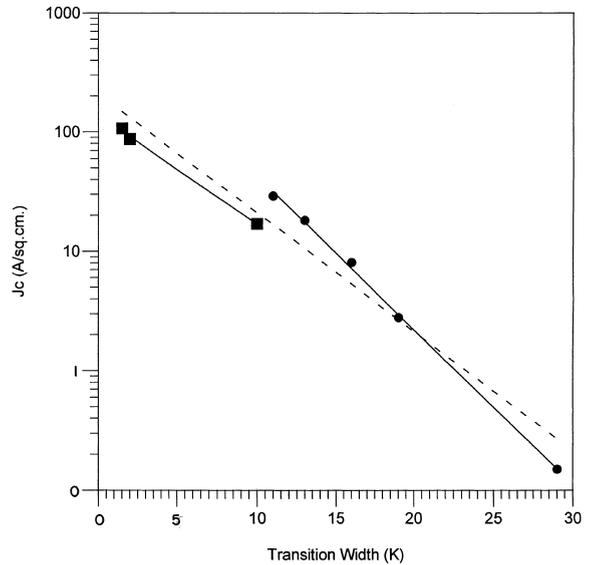

Fig. 6. The variation of the transport $J_c$ (measured at 77 K, $B = 0$) with the transition width $\Delta T_c$ in BPSCCO (●) and YBCO (■) samples.

Fig. 6. The pattern follows an exponential trend which is shown to be derivable using a simple heuristic argument. The transition width is found to be relatively smaller in the case of silver added samples which, again, reflects a certain degree of homogeneity within the matrix. The junction type also changes from superconductor–insulator–superconductor (SIS) to SNS as a result of silver addition. The transport critical current, therefore, improves in the silver added samples. It appears that the relations among the three important parameters, $J_c$, $\Delta T_c$, and $U$ are incompatible with each other – small $\Delta T_c$ leads to high $J_c$ yet small $U$, whereas high $\Delta T_c$ yields low $J_c$ though large $U$. In order to achieve high $J_c$ and $U$ yet small $\Delta T_c$, a balance is to be struck among these three parameters and an optimum level for each of these parameters is to be achieved.

## 4. Evaluation of the distribution function $m(U)$ for the flux pinning energy U

The flux pinning energy $U$ within a granular medium depends on the variation in the Josephson coupling energy $E_J$ ($\sim hI_c/2e$; $I_c =$ intergranular critical current) across the entire matrix. For an ordered two-dimensional junction network, where each junction is having identical junction coupling energy $E_J$, the flux pinning energy or the barrier height $U$ is calculated to be $0.2E_J$ [5]. For a disordered network, where $E_J$ varies from junction to junction, $U$ could be much higher which may lead to localization of the vortices. In a bulk granular medium, a distribution of the Josephson coupling energy $E_J$ (and hence a distribution of the flux pinning energy $U$ is expected as there is a large variation in the microstructural properties across the entire region. It results from the wide distribution of the particle size of starting powder, variation in the processing temperature from region to region, variation in the porosity, inadequate compaction pressure etc. It is possible to assume a certain distribution function for the flux pinning energy and calculate the magnetic relaxation pattern. Instead, we followed an inversion scheme [18] to evaluate the distribution function $m(U)$ from the experimentally observed magnetic relaxation of the intergranular critical state. Since, the intergranular magnetic moment $M(t,T)$ follows Anderson–Kim logarithmic decay pattern, it is given by

$$M(t,T) = M_0(T)\left[1 - \left\{\frac{k_B T}{U(T)}\right\}\ln\left(1 + \frac{t}{\tau}\right)\right], \quad (1)$$

where $\tau$ is the characteristic relaxation time, and $M_0$, the intergranular magnetic moment at time $t = 0$. Considering the distribution function $m(U)$ for the flux pinning energy $U$ and the functions $a(t')$ and $b(t')$ (where $t' = T/T_c$) for describing the temperature dependences of $M_0(T)$ and $U(T)$, the overall intergranular magnetic moment $M(t,T)$ can be written as

$$M(t,T) = M_0\left[\frac{b(t')}{a(t')}\right]\int_{U_0}^{\infty} m(U)\left[1 - \frac{k_B T}{U_0 b(t')}\ln\left(1 + \frac{t}{\tau}\right)\right]dU, \quad (2)$$

where $m(U)$ defines the fraction of the entire matrix whose pinning energy lies within the range $U$ and $U + dU$. $m(U)$ satisfies the condition,

$$\int_0^{\infty} m(U)dU = 1.$$

The average flux pinning energy $\langle U \rangle$ is given by $\langle U \rangle = \int U m(U) dU$. The functions $a(t')$ and $b(t')$ are related to the temperature dependence of $E_J(T)$ and takes the form $\sim (1 - t')^n$, where $n$ is expected to vary between 1.0 and 2.0 depending on the type of Josephson junctions SIS or SNS. $U(T)$ is given by $b^{-1}(t')[k_B T \ln\{1 + (t/\tau)\}]$ [18]. In contrast to the intragranular case, the functions $a(t')$ and $b(t')$ will have identical temperature dependence as both are related to $E_J(T)$. In order to evaluate $m(U)$ from Eq. (2), we need to invert the expression. Then, the expression for $m(U)$ is

$$m(U) = \left[\frac{d}{dT}\left\{\frac{a(t')}{M_0 k_B T}\right\}\frac{dM}{d\ln t}\right]\left[\frac{b(t')}{T}\frac{d}{dT}\left\{\frac{T}{b(t')}\right\}\right]^{-1}. \quad (3)$$

In order to completely evaluate $m(U)$ we have to calculate $\ln(t/\tau)$ initially which can be expressed as

$$\ln\left(\frac{t}{\tau}\right) = \left[\left\{T\frac{b(t')}{a(t')}\right\}\frac{d}{dT}\left\{M(t,T)\frac{a(t')}{b(t')}\right\}\right] \times \left[\left\{\frac{dM}{d\ln t}\right\}\left\{1 - \frac{d\ln b(t')}{d\ln T}\right\}\right]^{-1} \quad (4)$$

From the experimentally observed $M(t,T)$, where $t$ can be taken as a specific time $t_b$ ($= 600$ s in our case), and $dM(t,T)/d\ln t$, we can calculate first $\ln(t/\tau)$. The nature of the functions $a(t')$ and $b(t')$, i.e., the indices $n$ for each of the samples are listed in Table 1. These indices are the fitting parameters which are adjusted in order to arrive at near constancy in $\tau$. For the chosen set of parameters, the temperature dependency of $\tau$ is found to be minimum, though it decreases a little at higher temperature. The average value of $\tau$ for each of the samples is also listed in Table 1. The values of $n$ in the case of silver added samples are consistent with the range observed in SNS junctions. Hence, it reflects the presence of SNS junctions in silver added samples. Finally, by using Eq. (3), we have extracted the distribution function $m(U)$ for all the samples. Representative patterns are shown in Fig. 7. Notable is the change in pattern in the case of silver added samples for both YBCO as well as BPSCCO systems. Whereas the distribution is wider in the parent sample, it is narrower in the silver-added case. The pinning energy $U_m$ corresponding to the peak in the distribution function is shifted also from higher to lower side in the case of silver-added samples. The variance $\sigma^2$ provides a good measure of the extent of distribution. The ratio $\sigma_s^2/\sigma_p^2$ (where $\sigma_p^2$ and $\sigma_s^2$ define the variances of the distributions in the parent ceramic and silver added samples, respectively) helps in quantitatively estimating the degree of disorder within the entire granular medium. The variation of the ratio $\sigma_s^2/\sigma_p^2$ with the amount of silver addition ($x$ wt.%) is plotted in Fig. 8. Magnetic relaxation data in Ref. [4], corresponding to 2 and 4 wt.% silver-added samples, were also used for estimating the ratio $\sigma_s^2/\sigma_p^2$. Although within our chosen limit, the trend is downwards, i.e., the degree of disorder is decreasing with the increase in the amount of silver addition, with much higher silver the disorder can again rise as it will lead to inhomogeneous distribution of excess silver.

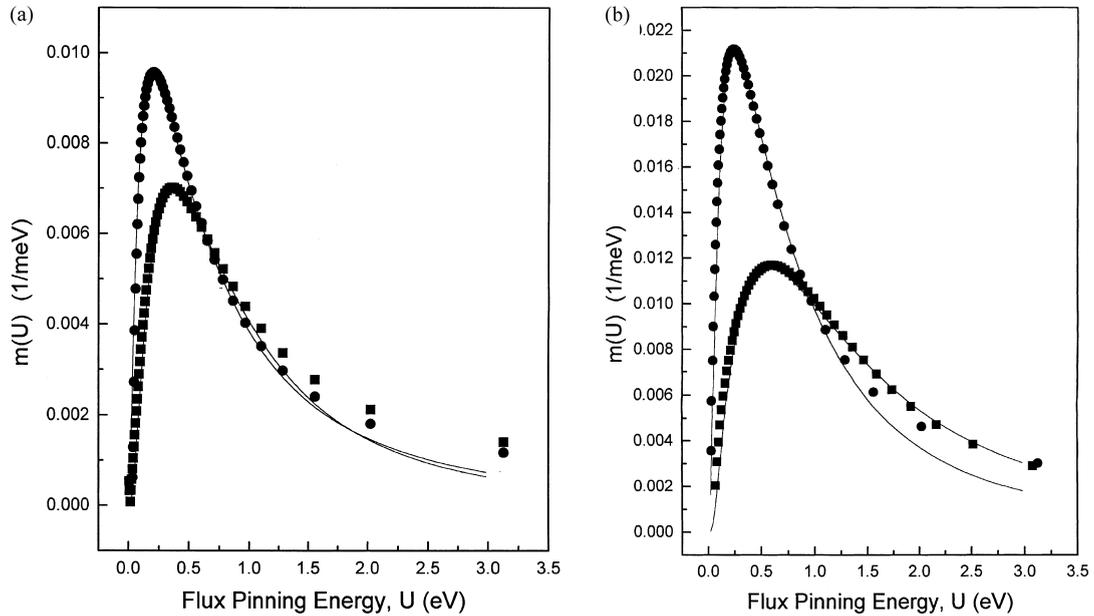

Fig. 7. The pattern of the distribution function for the flux pinning energy $m(U)$ for (a) BPSCCO (■), BPSCCO + 10 wt.% Ag (●) and (b) YBCO (■), YBCO + 15 wt.% Ag (●). The solid lines represent the fit with a log-normal function.

The pattern of the distribution can be fitted with log-normal function $p(U) = p_m \exp[-\gamma\{\ln(U/U_m)\}^2]$ in all the cases; $p_m$ defines the amount of fraction corresponding to the peak of the pattern and $U_m$ is the corresponding value of the pinning energy. The reason behind such log-normal type distribution of the intergranular flux pinning energy could be ascertained from the fact that the distribution of the grain boundary features is found to follow similar log-normal pattern [19]. The solid lines in Fig. 7 are the fits with the log-normal pattern. The fitting parameters $p_m$, $\gamma$ and $U_m$ are listed in Table 1.

We have also calculated the effective mass $m^*$ of the vortices by equating the kinetic energy with the change in potential energy. The net potential energy barrier is the summation of pinning energy and the intervortex repulsion energy for a finite concentration of the vortices. The intervortex repulsion energy in the continuum case is given by $\Delta F = (\Phi_0^2/8\pi^2\lambda^2)\ln(\lambda/r)$, if the intervortex separation $r \sim \lambda$. Replacing $\lambda$ by the expression $\lambda_\perp = \Phi_0^2 c/8\pi^2 E_J$ valid for the granular case [5], we can express the intervortex repulsion energy for the granular case. Obviously, the movement of the flux lines under a current flow will lead to the re-adjustment of the intervortex separation and the repulsion energy. The equilibrium state will have the minimum energy. Because of the flux line movement the coupling energy will vary and the equilibrium value $E_{Jm}$ can be obtained by minimizing the expression for the intervortex repulsion energy. The final expression of the minimum intervortex energy is $\Delta F_{min} = \Phi_0^2/16\pi^2 er^2$, where $r^2 \approx (\Phi_0/B)$ and is constant for a fixed $B$. By treating the flux lines as classical particles, the estimation of the effective mass $m^*$ of the vortices can be attempted using the expression,

$$(1/2)m^* v_d^2 fp^2 = \langle U \rangle + \Delta F_{min}, \quad (5)$$

where $fp^2$ is the number of the flux lines under a certain applied field, $v_d$ is the drift velocity of the vortices. $\langle U \rangle$ can be evaluated from $\langle U \rangle = \int U m(U) dU$,[2] where $m(U)$ is the log-normal function. Obviously, the higher is the pinning energy higher is the effective vortex mass. For an ordered two-dimensional array, the flux pinning energy is $\sim 0.2 E_J$, whereas the total energy of the core plaquatte is $4 E_J$. Therefore, only 5% of the total energy variation leads to the development of the flux pinning barrier. In the case of a disordered array, on the other hand, the order of variation in the energy of the core plaquatte could be as high as $\sim 100\%$ which leads to much higher flux pinning barrier. Using the relevant expression of $m(U)$ and right set of parameters, such as $\gamma$, $p_m$ and $U_m$ for the parent and the silver added samples and considering $v_d$ and $B$ to be same in both the cases we can calculate the ratio of the effective masses of the vortices. The effective mass $m^*$ for the silver-added samples is found to be $\sim 22\%$ of the values in the parent samples in the case of BPSCCO samples while $\sim 18\%$ in the case of YBCO samples. It is to be noted that the effective mass $m^*$ here is related directly to the interaction through the potential hill structure in entire space and *not* to the localization–delocalization phenomenon in any single junction. In the later case, the mass of a vortex $m_v$ is given by $m_v = \Phi_0^2 C/2a^2$, where $C$ is junction capacitance and $a$, the lattice parameter. It has already been noted that the vortex mass is zero in a SNS junction while it is finite in SIS junction [20]. Therefore, it is possible to reduce the effective vortex mass considerably by suitably controlling the amount of silver addition.

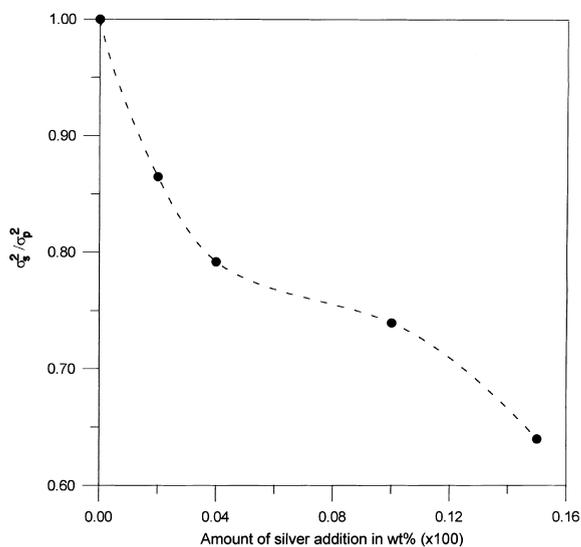

Fig. 8. The plot of the normalized variance $\sigma_s^2/\sigma_p^2$ of the distribution function $m(U)$, where $\sigma_p^2$ and $\sigma_s^2$ define the variances corresponding to the parent and the silver added samples, vs. the amount of silver addition (wt.%).

## 5. Evaluation of the distribution function $n(\theta)$ for the grain boundary misalignment angle $\theta$

We have developed a scheme for evaluating the distribution of the grain boundary misalignment angle $\theta$ from the observed pattern of variation in the critical fields of the grain boundaries with temperature. We assume that the grain dimensions in the systems do not vary by a large extent while the grain boundary misalignment angle varies substantially, leading to the formation of strongly coupled grain colonies as well as colonies of weaker grain boundaries across the bulk of the sample. Such distribution of weakly coupled and strongly coupled grain colonies in polycrystalline systems has been observed in recent magneto-optical study [21]. The impact of such variation on the overall transport critical current is enormous.

---

[2] Since $m(U) = p_m \exp[-\gamma \ln(U/U_m)^2]$, $\langle U \rangle = (1/2)p_m U_m^2 \sqrt{\pi/\gamma} \exp(1/\gamma)$ which allows exact calculation of the average flux pinning energy using the evaluated parameters $p_m$, $U_m$ and $\gamma$.

It also governs the critical fields of the grain boundaries. We show here that with a simple scheme one can extract the distribution of the grain boundary misalignment angle.

From the initial magnetization curves (Fig. 2), we have noted the lower critical field $H_{c1j}(T)$ at a given temperature (which, of course, corresponds to the full penetration field since at such a field the flux penetration at the grain boundaries throughout the entire matrix is almost complete). The temperature variation of $H_{c1j}$ is shown in Fig. 5. For a granular medium, the lower critical field of a grain boundary is given by [7]

$$H_{c1j}(T) = \frac{\mu_0 I_c}{8\pi a} \ln\left(\frac{3\pi^2 \phi_0}{4\mu_0 I_c a}\right), \quad (6)$$

which can be derived from the relation $H_{c1j} = (\Phi_0/4\pi\lambda_J^2)\ln(\lambda_J/\xi_J)$ by using suitable relations for the penetration depth $\lambda_J$ and coherence length $\xi_J$ valid in the case of granular medium: $\lambda_J = (\Phi_0 a/\mu_0 I_c)^{1/2}$, $\xi_J = 2a/\sqrt{3\pi}$ [7]. $I_c$ is the intergranular critical current and $a$ is the junction length. For perfectly aligned grains, the junction dimensions are considered to be length $= a$, width $= a$ and thickness $= t_{d0}$ (Fig. 9). For two misaligned grains having a misalignment angle $\theta$, the thickness $t_d$ varies across the junction dimensions. Considering all other junction dimensions (i.e., the length and the width) are remaining same, it is possible to show that the equivalent junction thickness is given by $t_d = t_{d0} + (1/2)a\tan\theta$ (Fig. 9). The junction normal-state resistance $R_n$, therefore, is given by $R_n = \rho_{n\square}[t_{d0} + (1/2)a\tan\theta]$ where $\rho_{n\square}$ is

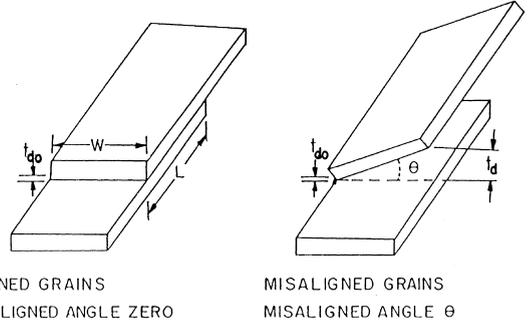

Fig. 9. The idealized grain boundary for two well-aligned grains with zero misalignment angle and with a misalignment angle $\theta$. If the other junction dimensions like length and width are considered as remaining same, the junction thickness ($t_d$) varies with the misalignment angle following $t_d = t_{d0} + (1/2)a\tan\theta$, where $t_{d0}$ is the junction barrier thickness for well-aligned grains and $a$ is the junction length and width.

work of Dimos et al. [22] that the junction $J_c$ scales as $\sim 1/\theta$ which implies that the junction resistance should scale as $\sim \theta$. This is automatically observed in our assumed relation between the junction resistance and the misalignment angle $\theta$. With the increase in grain boundary misalignment angle $\theta$, the junction normal-state resistance will increase which, in turn, will reduce the junction critical current $I_c(T)$ as $I_c(T) = \pi\Delta^2(T)/4ek_B TR_n$ [23,24]; $\Delta(T)$ is the BCS energy gap. This inverse relationship between $I_c$ and $R_n$ is found to be valid for different types of junctions – SIS, SNS, point contacts, etc. [1]. Substituting $I_c$ in Eq. (6) by this expression and noting the relation between $R_n$ and $\theta$, we can express $H_{c1j}(T)$ as

$$H_{c1j}(T) = \frac{\mu_0 \Delta^2(T)}{32 a k_B T e \rho_{n0}\{t_{d0} + \frac{1}{2}a\tan\theta\}} \ln\left[\frac{3\pi\phi_0 e k_B T \rho_{n0}\{t_{d0} + \frac{1}{2}a\tan\theta\}}{\mu_0 \Delta^2(T) a}\right]. \quad (7)$$

the junction normal-state resistivity per square of the junction cross-sectional area. It is to be noted that for proximity junctions the grain boundary resistance is considered to be varying exponentially with the barrier thickness. However, we have considered a straightforward linear variation valid for different types of junctions. This consideration is justified as it has been observed in the seminal

With the increase in $\theta$, the penetration depth increases reflecting a poor shielding effect while $H_{c1j}$ decreases. In a disordered grain boundary network, there is a distribution of the grain boundary misalignment angle $\theta$ all through the matrix. Let $n(\theta)$ be the distribution function which signifies the fraction of the entire matrix lying within a range $\theta$ and $\theta + d\theta$. $n(\theta)$ should be nor-malized as $\int_0^\infty n(\theta)d\theta = 1.0$. As a result of such distribution, the experimentally observed $H_{c1j}$ reflects only an average of the entire pattern of distribution of $H_{c1j}(\theta, T)$, i.e., $\langle H_{c1j}(T)\rangle = \int H_{c1j}(\theta, T)n(\theta)d\theta$. With the help of the experimentally observed $\langle H_{c1j}(T)\rangle$ and the Eq. (7), it is possible to evaluate the distribution function $n(\theta)$. The final expression in terms of $n(\theta)$ is

$$n(\theta) = \left[\frac{d}{dT}\{H_{c1j}(T)\}\right]\left[\frac{A(T)}{t_{d0} + \frac{1}{2}a\tan\theta}\ln\left\{\frac{3\pi\phi_0(t_{d0} + \frac{1}{2}a\tan\theta)}{32a^2 A(T)}\right\}\frac{d\theta}{dT}\right]^{-1}, \qquad (8)$$

where $A(T) = \mu_0\Delta^2(T)/32aek_BT\rho_{n0}$; $\theta$ can be evaluated from the relation between $\theta$ and $R_n$ and hence between $\theta$ and $I_c(T)$ while $\rho_{n0}$ is considered to be $10^6$ ohm/m, $t_{d0} = 10^{-8}$ m and $a = 10^{-6}$ m. The representative patterns of $n(\theta)$ are shown in Fig. 10 for the parent and silver added BPSCCO samples. In the case of silver added samples, the distribution pattern is narrower with the peak shifted towards a smaller angle. Therefore, it also reflects that the microstructural parameters have become a bit more uniform with the addition of silver. A large fraction of the matrix, of course, lies within a higher angle regime which probably highlights the general trend in all such ceramic polycrystalline samples; only in the case of textured samples does the average grain-boundary misalignment angle shift toward much lower values. Silver addition, of course, does improve the microstructure as well as develops a uniformity all through the matrix. The distribution patterns can be approximately fitted with the log-normal function (Fig. 10; solid lines) using following set of parameter values: $p_m = 0.0098$, $\theta_m = 28.0$, $\gamma = 7.0$ for the silver added sample; $p_m = 0.008$, $\theta_m = 32.0$, $\gamma = 1.5$ for the parent sample. The change in the distribution pattern for the silver added case is not as prominent as it is in the case of distribution of flux pinning energy $U$.

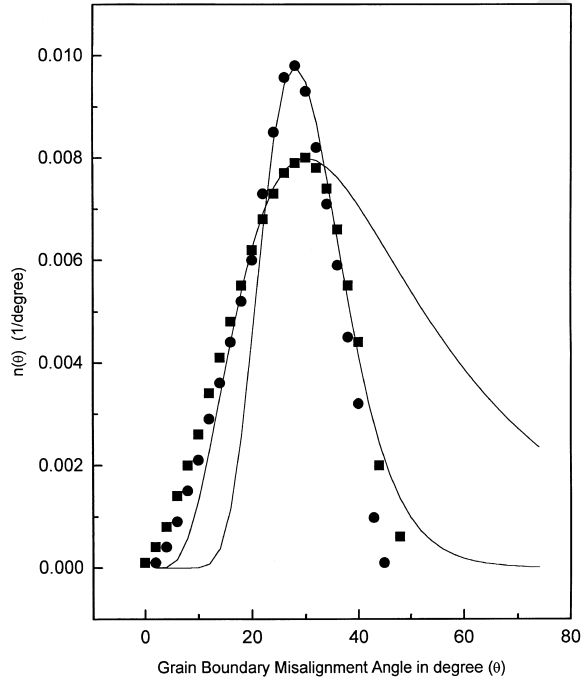

Fig. 10. The pattern of the distribution function for the grain boundary misalignment angle $n(\theta)$ for BPSCCO (■) and BPSCCO + 10 wt.% Ag (●).

## 6. Correlation among $J_c$, $U$ and $\Delta T_c$

We have noted a correlation among the three important parameters $J_c$, $U$ and $\Delta T_c$ which helps in quantitative estimation of the degree of disorder within the sample. In a polycrystalline sample, the macroscopic superconducting transition takes place at a certain temperature $T_c(R = 0)$ where the phase locking develops among the superconductor islands across the junction. For a two-dimensional array of junctions, such transition temperature is given by $T_c(R = 0) = (\pi/2)E_J/k_B$ [25] which corresponds to the vortex–antivortex pairing temperature. $T_{c,\text{onset}}$, on the other hand, corresponds to the intragranular $T_c$ and, therefore, characteristic of the superconducting composition. The transition width, $\Delta T_c$, reveals the nature of grain to grain coupling within the ensemble of grains through which the measuring current flows. The weakest link in the chain determines the transition width. For a certain measuring current (which, in our case, is 1 mA), the transition width will be very small if the grain to grain coupling all through the ensemble is strong (in other words, if $E_J$ is high and there is virtually no variation in $E_J$ from coupling to coupling) and vice versa. Hence, the transition width is roughly a measure of the degree of inhomogeneity within the sample. For higher measuring current $J$) and/or higher applied magnetic field ($H$), $\Delta T_c(H)$ becomes higher and takes different shapes of variation with $H$ depending on the degree of variation in $E_J$ under $H$ or $J$ across the ensemble. We have observed, in the case of our samples, that the transport $J_c$ (measured at 77 K under zero applied field) is scaling with the transition width $\Delta T_c$ as $\sim \exp(-\Delta T_c)$ where the $\Delta T_c$ is measured with a small current under zero applied field. Although, within a granular medium a relation between $J_c$ and $\Delta T_c$ is expected intuitively, such a formal relation helps in quantitative estimation of the effect of change in $\Delta T_c$ on $J_c$. It is possible to show that one can arrive at such an expression using a heuristic argument regarding transition within a granular network. The transition temperature $T_c(R=0)$ can be given by

$$k_B T_c(R=0) = \pi E_{J0} \ln(r_2/r_1), \tag{9}$$

where the right-hand side of the equation depicts the total Josephson coupling energy of an ordered two-dimensional granular medium when the vortex–antivortex pairing takes place [5]; $E_{J0}$ is the equilibrium value of the junction coupling energy for each junction corresponding to the transition temperature $T_c(R=0)$; $r_1$ and $r_2$ define the geometrical boundary of the medium. For a disordered network with a wide variation in the junction coupling energy $E_J$, the net flux pinning energy $U$ is given by the difference in the junction coupling energy from junction to junction. From Eq. (9), it is clear that such local variation in the junction coupling energy will lead to a corresponding local variation in the transition temperature as well across the entire network. Considering $E_{J0} + dE_J$ as the junction coupling energy corresponding to the transition temperature $T_c(R=0) + dT_c$, it is possible to show using Eq. (9) that the overall flux pinning energy $U$ is scaling linearly with the transition width $\Delta T_c$, where $\Delta T_c$ signifies the difference between the maximum and minimum transition temperature across the network:

$$U = \left\{ \frac{E_{J0}(T)}{T_c(R=0)} \right\} \Delta T_c. \tag{10}$$

Such a relationship can be noticed in the data of Norling et al. ([13] and references therein) [1] as well. It is to be noted that the $U$ here is the $J$-independent intrinsic pinning energy as the relation between $U$ and $\Delta T_c$ is calculated under static vortex scenario. Hence, $U$ here is intrinsically related to the uniformity or non-uniformity aspect across the disordered Josephson junction array. The $J$-dependent apparent $U(J)$ follows linear or non-linear pattern of variation yet reflects the signature of the intrinsic $U$ in both the silver-free and silver added cases. The apparent $U$ values calculated from the relaxation patterns (Section 3) are, therefore, related to the intrinsic pinning energy and their difference in silver-free and silver added samples can be related to the extent of uniformity of the junction matrix. We now use the Ambegaokar–Halperin relationship [26] for the grain boundary resistance $R/R_n = [I_0(U/2k_BT)]^{-2}$, where $I_0$ is the modified Bessel function. This relation is derived by considering the vortex kinetics in a current driven overdamped junction. We apply this relation in our case by considering $R_n$ as the maximum junction normal-state resistance and $R$ as the junction resistance for other junctions. Replacing $U$ by Eq. (10), we obtain

$$\frac{R}{R_n} = \left[ I_0 \left\{ K(T) \frac{\Delta T_c}{T_c(R=0)} \right\} \right]^{-2}, \tag{11}$$

where $K(T) = E_{J0}(T)/2k_BT$, a dimensionless yet temperature dependent parameter. Noting the Ambegaokar–Baratoff relationship [23,24] between the grain boundary resistance and the junction critical current and observing the fact that the Bessel function can be approximated to $\sim \exp\{-K(T)\Delta T_c/T_c(R=0)\}$ for large argument, we can write an approximate relation between $J_c$ and $\Delta T_c$ as

$$J_{c,\min}(T) \cong J_{c,\max}(T) \exp\left[-K(T)\frac{\Delta T_c}{T_c(R=0)}\right], \tag{12}$$

where $J_{c,\min}$ and $J_{c,\max}$ correspond to the junction resistances $R_n$ and $R$. If the argument of the Bessel function is less than one then the relationship will be a quadratic one. Therefore, depending on the values of $\Delta T_c$ one can expect a cross-over from exponential decay of the critical current with the transition width to a quadratic degradation. $R_n$ measures the maximum normal state resistance corresponding to the minimum junction critical current over the entire matrix and grows exponentially with the transition width. These quantitative relationships among $J_c$, $U$ and $\Delta T_c$ may help in suitably controlling the quality of the granular samples. It is interesting to note that the experimental results (Fig. 6) of variation of transport $J_c$ with $\Delta T_c$ follow exponential patterns with different set of parameters for YBCO and BPSCCO samples. In the case of the YBCO samples, the pattern can be fitted with Eq. (12) for the choice of the parameters: $J_{c,\max} \approx 140$ A/cm$^2$, $K(T)/T_c(R=0) \approx 0.21$, whereas in the case of the BPSCCO samples, the pattern can be fitted for $J_{c,\max} \approx 830$ A/cm$^2$, $K(T)/T_c(R=0) \approx 0.296$. It shows that the equilibrium coupling energy $E_{J0}(T)$ is higher in the BSCCO sample, probably, because it is less granular than the YBCO system. However, a crossover from exponential decay to a quadratic decay of $J_c$ is expected for lower values of $\Delta T_c$.

The apparent contradictory relation between the transport $J_c$ and the flux pinning energy $U$ within such Josephson junction network can be understood from the fact that the transport $J_c$ in such a medium depends strongly on the net current carrying cross-section rather than on the flux creeping effect. In silver added samples, large homogeneous areas with high current carrying capability are available which lead to high transport $J_c$. The impact of local microstructural inhomogeneity in modulating the transport $J_c$ has been observed directly by magneto-optical imaging [27]. The flux pinning energy $U$, on the other hand, depends on the degree of inhomogeneity. Therefore, though rise in inhomogeneity yields high $U$, it automatically leads to a drop in the current carrying cross-section and hence a drop in transport $J_c$. Such a strong correlation between $U$ and the current carrying cross-section is normally not observed in intragranular (or single crystal) case.

## 7. Summary and conclusions

In summary, ceramic bulk YBCO and BPSCCO superconductors have been prepared. Silver is added in such a matrix by controlled amount. Silver is found to have given rise to a grain boundary network with much uniform characteristics. The distribution function for the flux pinning energy $m(U)$ is extracted from the experimentally observed magnetic relaxation patterns. The distribution function for the grain boundary misalignment angle $n(\theta)$ is also evaluated by suitably inverting the experimentally observed pattern of variation of the grain boundary critical fields with temperature. Both the distribution functions depict a change in the pattern in the case of silver added samples. The distribution is narrower with the shift in the peak toward lower side. The three important parameters – $J_c$, $\Delta T_c$, and $U$ – are found to be interrelated: $J_c$ degrades exponentially with the increase in $\Delta T_c$ while $U$ is found to scale linearly with $\Delta T_c$. Such an apparent contradictory relation between $J_c$ and $U$ is the result of strong correlation between $U$ and overall current carrying cross-section: rise in $U$ leads to drop in current carrying cross-section. The quantitative relations among $J_c$, $U$ and $\Delta T_c$ will help in devising a strategy toward modulating the microstructural properties in order to achieve the desired effect: high $J_c$ and small decay rate.

## Acknowledgements

Authors are pleased to acknowledge the assistance rendered to them by Dr. A. Sen of Central Glass and Ceramic Research Institute (CGCRI), Calcutta, during the magnetic and microstructural studies which were carried out at CGCRI. They also wish to acknowledge helpful discussion with Prof. A.K. Raychaudhuri of NPL. One of the authors' (A.P.) thankfully acknowledges the fi-nancial support of UGC in the form of a research fellowship during the work.

## References

[1] M. Prester, Supercond. Sci. Technol. 11 (1998) 333.
[2] S.A. Sergeenkov, Physica C 205 (1993) 1.
[3] G. Costabile, et al. (Eds.), Nonlinear Superconducting Electronics and Josephson Devices, Plenum Press, New York, 1991.
[4] J.R. Kirtley, P. Chaudhari, M.B. Ketchen, N. Khare, S.Y. Lin, T. Shaw, Phys. Rev. B 51 (1995) 12057.
[5] C.J. Lobb, D.W. Abraham, M. Tinkham, Phys. Rev. B 27 (1983) 150.
[6] A.H. Tuohimaa, J.A.J. Paasi, Physica C 319 (1999) 73.
[7] M. Tinkham, C.J. Lobb, in: H. Ehreinrich, D. Turnbull (Eds.), Solid State Physics, vol. 42, Academic Press, New York, 1989, p. 91.
[8] R. Fehrenbacher, V.B. Geshkenbein, G. Blatter, Phys. Rev. B 45 (1992) 5450.
[9] M.A. Itzler, M. Tinkham, Phys. Rev. B 51 (1995) 435.
[10] M.A. Itzler, M. Tinkham, Phys. Rev. B 53 (1996) R11949.
[11] M.A.-K. Mohamed, I. Isaac, J. Jung, Physica C 235–240 (1994) 3333.
[12] J. Jung, M.A.-K. Mohamed, I. Isaac, L. Freidrich, Phys. Rev. B 49 (1994) 12188.
[13] P. Norling, K. Niskanen, J. Magnusson, P. Nordblad, P. Svedlindh, Physica C 221 (1994) 169.
[14] A. Pandey, Y.S. Reddy, R.G. Sharma, J. Mater. Sci. 32 (1997) 3701.
[15] P. Sujatha Devi, H.S. Maiti, J. Solid State Chem. 109 (1994) 35.
[16] S. Bernik, Supercond. Sci. Technol. 10 (1997) 671.
[17] Y. Yeshurun, A.P. Malozemoff, A. Shaulov, Rev. Mod. Phys. 68 (1996) 911.
[18] C.W. Hagen, R. Griessen, Phys. Rev. Lett. 62 (1989) 2857.
[19] C.G. Granqvist, R.A. Buhrman, J. Appl. Phys. 47 (1976) 2200.
[20] F. Gibson, A.T. Gongora, J.V. Jose, Phys. Rev. B 58 (1998) 982.
[21] U. Welp, D.O. Gunter, G.W. Crabtree, W. Zhang, U. Balachandran, P. Halder, R.S. Sokolowski, V.K. Vlasko-Vlasov, V.I. Nikitenko, Nature 376 (1995) 44.
[22] D. Dimos, P. Chaudhari, J. Mannhart, F.K. LeGoues, Phys. Rev. Lett. 61 (1988) 219.
[23] V. Ambegaokar, A. Baratoff, Phys. Rev. Lett. 10 (1963) 486.
[24] V. Ambegaokar, A. Baratoff, Phys. Rev. Lett. 11 (E) (1963) 104.
[25] J.M. Kosterlitz, D.J. Thouless, J. Phys. C 6 (1973) 1181.
[26] V. Ambegaokar, B.I. Halperin, Phys. Rev. Lett. 22 (1969) 1364.
[27] A.E. Pashitski, A. Gurevich, A.A. Polyanskii, D.C. Larbalestier, A. Goyal, E.D. Specht, D.M. Kroeger, J.A. DeLuca, J.E. Tkaczyk, Science 275 (1997) 367.